\documentclass[
  ,draft            
  ,numberedheadings 
  ]
  {aipproc}

\layoutstyle{6x9}

\begin{document}

\title{The Standard-smooth Variant of Hybrid 
Inflation}


\author{George Lazarides}{
address={Physics Division, School of Technology, 
Aristotle University of Thessaloniki,
Thessaloniki 54124,
Greece}
}

\begin{abstract}

We consider the extension of the supersymmetric 
Pati-Salam model introduced in order to solve the 
$b$-quark mass problem in supersymmetric theories 
with Yukawa unification, universal boundary 
conditions and $\mu>0$. This model naturally 
leads to the new shifted and new smooth hybrid 
inflation scenarios, which, however, yield, in 
minimal supergravity, too large values of the 
spectral index $n_{\rm s}$. We show that this problem 
can also be resolved within the same model by a 
two-stage inflationary scenario based only on 
renormalizable superpotential interactions. The 
first stage is of the standard and the second of 
the new smooth hybrid type.
The cosmological scales exit the horizon during 
the first stage of inflation and acceptable $n_{\rm s}$'s 
can be achieved by restricting the number of 
e-foldings of our present horizon during this 
inflationary stage. The additional
e-foldings needed for solving the horizon and
flatness problems are naturally provided by the
second stage of inflation.
Monopoles are formed at the end of the first
stage of inflation and are, subsequently, diluted
by the second stage of inflation so that their 
density in the present universe is utterly 
negligible.
 
\end{abstract}

\maketitle

\section{Introduction}

One of the most promising inflationary scenarios is 
hybrid inflation (HI) \cite{linde}, which 
is naturally realized \cite{cop,rc} in 
supersymmetric (SUSY) grand unified theory (GUT) models. 
In its standard realization, though, the GUT gauge 
symmetry ($G$) breaking occurs at the end of inflation 
leading \cite{smooth1} to a disastrous 
production of magnetic monopoles if these defects are 
predicted by the underlying symmetry breaking. This 
problem is avoided in the smooth \cite{smooth1,smooth2} 
or shifted \cite{shift} variants of SUSY HI, where $G$ 
is broken to the standard model gauge group 
($G_{\rm SM}$) already during inflation (for a review, 
see Ref.~\cite{talks}).

These two variants of SUSY HI, originally based on 
non-renormalizable superpotential terms, can be
implemented \cite{nshift,nsmooth} with only 
renormalizable terms in an extended SUSY GUT model 
based on the Pati-Salam (PS) gauge group 
$G_{\rm PS}={\rm SU}(4)_c\times {\rm SU}(2)_{\rm L}
\times{\rm SU}(2)_{\rm R}$ \cite{pati}, whose breaking 
to $G_{\rm SM}$ predicts \cite{magg} the existence of 
doubly charged monopoles.

It is very interesting to point out that this 
extended SUSY PS model was motivated \cite{quasi}
(see also Ref.~\cite{quasitalks}) by a very
different issue: In SUSY models with Yukawa 
unification (YU)\cite{als} , such as the simplest 
SUSY PS model (see Ref.~\cite{hw}), and universal 
boundary conditions, the mass of the $b$-quark 
($m_b$) turns out \cite{hall} to be too large 
for $\mu>0$. By appropriately extending the model, 
however, YU is modestly violated and $m_b$ can be 
adequately reduced.

Fitting the Wilkinson microwave anisotropy probe 
(WMAP) data \cite{wmap} with the standard 
power-law cosmological model with cold dark matter 
and a cosmological constant, one 
obtains a spectral index $n_{\rm s}$ clearly lower 
than unity. However, in canonical supergravity 
(SUGRA), these HI models yield \cite{senoguz} 
$n_{\rm s}$'s which are very close to unity or 
even larger than unity.

One way \cite{mhin} to reduce $n_{\rm s}$ is by 
restricting the number of e-foldings of our 
present horizon during the main part of inflation 
which is responsible for the observed density 
perturbations. The additional e-foldings required 
for solving the horizon and flatness problems can 
be provided by a subsequent second stage of 
inflation.

We show \cite{sshi} that the same extended SUSY PS 
model of Ref.~\cite{quasi} leads to a two-stage 
inflationary scenario with acceptable values of 
$n_{\rm s}$ in minimal SUGRA. The first stage is of 
the standard hybrid type, while the second is of 
the smooth hybrid type. So, the name standard-smooth 
HI is coined for this scenario. 
Standard HI occurs along a trivial flat 
direction on  which $G_{\rm PS}$ is unbroken. As the 
inflaton drops below a critical value, this direction 
is destabilized giving its place to a non-flat valley 
for smooth HI on which $G_{\rm PS}$ is broken.

Note that the same extended SUSY PS model can lead 
\cite{hics} to an alternative inflationary scenario
with cosmic strings \cite{string} (for a textbook
presentation or a review, see e.g. 
Ref.~\cite{vilenkin}), since it possesses a 
semi-shifted 
flat direction on which ${\rm U}(1)_{\rm B-L}$ is 
unbroken. When the system crosses the critical point 
of this path, ${\rm U}(1)_{\rm B-L}$ breaks and 
cosmic strings are produced contributing to the 
primordial perturbations, in which case, the data 
allow \cite{bevis1} larger values of $n_{\rm s}$. 
In this semi-shifted inflationary scenario, 
no magnetic monopoles are formed. 

\section{The extended SUSY PS Model}

In this model \cite{quasi} (see also
Refs.~\cite{shift,quasitalks}), the breaking of 
$G_{\rm PS}$ to $G_{\rm SM}$ is achieved by the 
vacuum expectation values (VEVs) of a conjugate pair 
of Higgs superfields $H^c$ and $\bar{H}^c$ belonging 
to the $(\bar{4},1,2)$ and $(4,1,2)$ representations 
of $G_{\rm PS}$ respectively. There also exist a 
gauge singlet $S$ and a conjugate pair of superfields 
$\phi$, $\bar{\phi}\in (15,1,3)$ with 
$\langle\phi\rangle$, the VEV of $\phi$, breaking 
$G_{\rm PS}$ to $G_{\rm SM}\times{\rm U}(1)_{\rm B-L}$.
In addition to $G_{\rm PS}$, the model possesses a $Z_2$ 
matter parity symmetry and two global ${\rm U}(1)$ 
symmetries, namely a Peccei-Quinn and a R symmetry.
 
The superpotential terms relevant for inflation are
\cite{nsmooth}
\begin{equation}
W=\kappa S(M^2-\phi^2)-\gamma S H^c\bar{H}^c+
m\phi\bar{\phi}-\lambda\bar{\phi} H^c\bar{H}^c,
\label{W}
\end{equation}
where the mass parameters $M$, $m$ ($\sim M_{\rm GUT}$, 
the SUSY GUT scale) and any two of the three 
dimensionless coupling constants $\kappa$, $\gamma$, 
$\lambda$ can be made real and positive by an 
appropriate rephasing of the fields. We choose the 
third dimensionless coupling constant to be real and 
positive too.

The F--term scalar potential obtained from the 
superpotential $W$ in Eq.~\eqref{W} is
\begin{eqnarray}	
V&=&|\kappa\,(M^2-\phi^2)-\gamma H^c\bar{H}^c|^2+
|m\bar{\phi}-2\kappa S\phi|^2+|m\phi-\lambda H^c\bar{H}^c|^2
\nonumber\\
& &+|\gamma S+\lambda\bar{\phi}\,|^2\left(|H^c|^2+|\bar{H}^c|^2
\right).
\label{V}
\end{eqnarray}

From this potential and the vanishing of the D--terms 
(which yields $\bar{H}^{c*}= e^{i\theta}H^c$), we find 
\cite{nsmooth} two distinct continua of SUSY vacua:
\begin{eqnarray}	
\phi=\phi_{+},\quad\bar{H}^{c*}=H^c,\quad |H^c|=
\sqrt{m\phi_{+}/\lambda} \quad (\theta=0);
\label{susyvacua1}
\\
\phi=\phi_{-},\quad\bar{H}^{c*}=-H^c,\quad |H^c|=
\sqrt{-m\phi_{-}/\lambda} \quad
(\theta=\pi)
\label{susyvacua2}
\end{eqnarray}
with $\bar{\phi}=S=0$, where
\begin{equation}
\phi_{\pm}\equiv\frac{\gamma m}{2\kappa\lambda}\left(
-1\pm\sqrt{1+\frac{4\kappa^2\lambda^2M^2}
{\gamma^2m^2}}\,\right).
\nonumber
\end{equation}

\section{Flat Directions}

The potential $V$ in Eq.~\eqref{V} possesses 
\cite{nsmooth} three flat directions:

$(i)$ The trivial flat direction: 
$\phi=\bar{\phi}=H^c=\bar{H}^c=0$ with 
$V=V^0_{\rm tr}\equiv\kappa^2M^4$.

$(ii)$ The new shifted one (for $\gamma\neq 0$)
on which $G_{\rm PS}$ is broken to $G_{\rm SM}$:
\begin{eqnarray}	
\begin{array}{rcl}
\phi=-\frac{\gamma m}{2\kappa\lambda},\quad
\bar{\phi}=-\frac{\gamma}{\lambda}\,S,\quad
H^c\bar{H}^c=\frac{\kappa\gamma(M^2-\phi^2)+\lambda m\phi}
{\gamma^2+\lambda^2},
\\
V=V^0_{\rm nsh}\equiv\frac{\kappa^2\lambda^2}
{\gamma^2+\lambda^2}\left(M^2+\frac{\gamma^2m^2}
{4\kappa^2\lambda^2}\right)^2.~~~~~~~~~~~~~
\end{array}
\label{newshiftedpath}
\end{eqnarray}	

$(iii)$ The semi-shifted one (for
$\tilde{M}^2\equiv M^2-m^2/2\kappa^2>0$):
\begin{equation}
\phi=\pm\,\tilde{M},\quad
\bar{\phi}=\frac{2\kappa\phi}{m}\,S,\quad
H^c=\bar{H}^c=0
\end{equation} 
on which $G_{\rm PS}$ is broken to
$G_{\rm SM}\times {\rm U(1)_{B-L}}$ ($\phi\neq 0$, 
$H^c=\bar{H}^c=0$) and
$V=V^0_{\rm ssh}\equiv\kappa^2(M^4-\tilde{M}^4)$.

\section{Standard-smooth HI}

We will take $\tilde{M}^2<0$.
In this case, the semi-shifted flat direction
does not exist. Also, $V^0_{\rm nsh}>V^0_{\rm tr}$
and, thus, the system will eventually settle down on
the trivial flat direction. Expanding $\phi$, 
$\bar{\phi}$, $H^c$, $\bar{H}^c$ as
$s=(s_1+i\,s_2)/\sqrt{2}$, we find \cite{sshi}, 
on the trivial direction, the
${\rm mass}^2$ matrices $M_{\phi1}^2$ of
$\phi_1$, $\bar{\phi}_1$ and $M_{\phi2}^2$ of
$\phi_2$, $\bar{\phi}_2$:
\begin{equation}
\label{Mphi}
M_{\phi1(\phi2)}^2=\left(\begin{array}{cc}
m^2+4\kappa^2|S|^2\mp2\kappa^2M^2 & -2\kappa m S \\
-2\kappa m S  & m^2 \end{array}\right)
\end{equation}
and the ${\rm mass}^2$ matrices $M_{H1}^2$ of
$H^c_1$, $\bar{H}^c_1$ and $M_{H2}^2$ of $H^c_2$,
$\bar{H}^c_2$:
\begin{equation}
\label{MH}
M_{H1(H2)}^{2}=\left(\begin{array}{cc}
\gamma^2|S|^2  & \mp\gamma\kappa M^2  \\
\mp\gamma\kappa M^2 & \gamma^2|S|^2 \end{array}\right).
\end{equation}
The matrices $M_{H1(H2)}^2$ acquire one negative 
eigenvalue for $|S|<S_c\equiv\sqrt{\kappa/\gamma}\;M$
and the trivial direction becomes unstable.

For $\gamma\ll 1$, the trivial direction,
after its destabilization at $S_c$, gives its place
\cite{nsmooth} to a valley of minima with 
$\theta\simeq 0$, which leads to the corresponding 
SUSY vacua in Eq.~\eqref{susyvacua1}. This valley 
possesses a classical inclination and can 
accommodate a second stage of inflation of the new 
smooth hybrid type.

The standard-smooth HI scenario
goes \cite{sshi} as follows:

$(i)$ The system initially inflates on the
trivial direction which acquires a slope from
the one-loop radiative correction (RC) \cite{rc} 
due to the SUSY breaking caused by the constant 
potential energy density $V=V^0_{\rm tr}$. So, we 
get a first stage of inflation of the standard 
hybrid type.

$(ii)$ As the system moves below $S_c$,
$G_{\rm PS}$ breaks to $G_{\rm SM}$. After a 
short intermediate inflationary phase, the 
system settles down on the new smooth path and
new smooth HI occurs. The second stage of 
inflation (intermediate phase plus new
smooth HI) yields the additional e-foldings for 
solving the horizon and flatness problems.
At the end, the system falls into the
SUSY vacua leading, though, to no magnetic 
monopoles, since $G_{\rm PS}$ is broken to 
$G_{\rm SM}$ during the second stage of 
inflation.

However, two important requirements must
be fulfilled \cite{sshi}:

$(a)$ The number of e-foldings during the
second stage of inflation must be adequately 
large for diluting any monopoles generated at 
the end of the first stage.

$(b)$ Cosmological scales get perturbations
only from the first stage of inflation.

\subsection{One-loop RC}

The one-loop RC \cite{rc} to the potential $V$ from 
SUSY breaking on the trivial path is calculated by 
the Coleman-Weinberg formula \cite{ColemanWeinberg}:
\begin{equation}
\Delta V=\frac{1}{64\pi^2}\;\sum_i(-1)^{F_i}M_i^4
\ln\frac{M_i^2}{\Lambda^2},
\end{equation}
where we sum over helicity states, $F_i$ and 
$M_i^2$ are the fermion number
and ${\rm mass}^2$ of the $i$th state and
$\Lambda$ is a renormalization scale.

So, we need the mass spectrum of the model 
on the trivial path. We find \cite{sshi} two 
groups of 45 pairs of real scalars with 
${\rm mass}^2$ matrices
\begin{equation}
M_{-(+)}^2=\left(\begin{array}{cc}
m^2+4\kappa^2|S|^2\mp2\kappa^2M^2 & -2\kappa m S \\
-2\kappa m S  & m^2 \end{array}\right)
\end{equation}
and two more groups of 8 pairs of real scalars
with ${\rm mass}^2$ matrices
\begin{equation}
M_{1(2)}^{2}=\left(\begin{array}{cc}
\gamma^2|S|^2  & \mp\gamma\kappa M^2  \\
\mp\gamma\kappa M^2 & \gamma^2|S|^2 \end{array}
\right).
\end{equation}
Also, 45 pairs of Weyl fermions with
${\rm mass}^2$ matrix
\begin{equation}
M_0^2=\left(\begin{array}{cc}
m^2+4\kappa^2|S|^2 & -2\kappa m S \\
-2\kappa m S  & m^2 \end{array}\right)
\end{equation}
and 8 more pairs of Weyl fermions with
${\rm mass}^2$ matrix
\begin{equation}
\bar{M}_0^2=\left(\begin{array}{cc}
\gamma^2|S|^2  & 0  \\
0 & \gamma^2|S|^2 \end{array}\right).
\end{equation}

The one-loop RC to the inflationary potential 
$V$ is then
\begin{eqnarray}
\Delta V &=&
\frac{45}{64\pi^2}{\rm tr}\left[
M_{+}^4\ln\frac{M_{+}^2}{\Lambda^2}+
M_{-}^4\ln\frac{M_{-}^2}{\Lambda^2}
-2M_{0}^4\ln\frac{M_{0}^2}{\Lambda^2}
\right]
\nonumber\\
& &+\frac{8}{64\pi^2}{\rm tr}\left[
M_{1}^4\ln\frac{M_{1}^2}{\Lambda^2}+
M_{2}^4\ln\frac{M_{2}^2}{\Lambda^2}-
2\bar{M}_{0}^4\ln\frac{\bar{M}_{0}^2}
{\Lambda^2}
\right].
\label{oneloop}
\end{eqnarray}
The total effective potential on the trivial 
path in global SUSY is
\begin{equation}
V_{\rm tr}=v_0^4+\Delta V,
\end{equation}
where $v_0\equiv\sqrt{\kappa}M$ is the
inflationary scale. Note that the 
$\sum_i(-1)^{F_i}M_i^4$ is $S$-independent implying 
that the slope of the path is
$\Lambda$-independent. This guarantees that the 
observables do not depend on $\Lambda$, which
remains undetermined.

\subsection{SUGRA correction}

The F--term scalar potential in SUGRA is
\begin{equation}
V=e^{K/m_{\rm P}^2}
\left[(F_i)^* K^{i^*j}F_j-3\,\frac{|W|^2}
{m_{\rm P}^2}\right],
\end{equation}
where $K$ is the K\"{a}hler potential,
$F_i=W_i+K_iW/m_{\rm P}^2$, a subscript $i$
($i^*$) denotes derivative with respect to 
the complex scalar $s^i$ ($s^{i{\,*}}$),
$K^{i^*j}$ is the inverse of $K_{j\,i^*}$,
and $m_{\rm P}$ is the reduced Planck mass.
We will only consider minimal K\"{a}hler
potential
\begin{equation}
K^{\rm min}=|S|^2+|\phi|^2+|\bar{\phi}|^2+|H^c|^2+
|\bar{H}^c|^2.
\end{equation}
The F--term scalar potential then becomes ($s$ is 
any of the five complex scalars above)
\begin{equation}
V^{\rm min}=e^{K^{\rm min}/m_{\rm P}^2}\;\left[
\sum_{s}\left|W_s+\frac{Ws^*}{m_{\rm P}^2}
\right|^2-3\,\frac{|W|^2}{m_{\rm P}^2}\right].
\end{equation}

On the trivial path, this scalar potential up to 
4th order in $|S|$ is \cite{sshi}
\begin{equation}
V^{\rm min}_{\rm tr}\simeq v_0^4\left
(1+\frac{1}{2}\,\frac{|S|^4}{m_{\rm P}^4}\right)
\end{equation}
and the effective potential for the standard
hybrid case becomes
\begin{equation}
\label{Vtrsugra}
V^{\rm SUGRA}_{\rm tr}\simeq V^{\rm min}_{\rm tr}
+\Delta V
\label{Vtr}
\end{equation}
with $\Delta V$ being the one-loop RC in 
Eq.~\eqref{oneloop}.

The effective potential on the new smooth path becomes
\cite{sshi}
\begin{equation}
V^{\rm SUGRA}_{\rm nsm}\simeq v_0^4\left(
\tilde{V}_{\rm nsm}
+\frac{1}{2}\,\frac{|S|^4}{m_{\rm P}^4}\right),
\label{Vnsm}
\end{equation}
where $\tilde{V}_{\rm nsm}\equiv
V_{\rm nsm}/v_0^4$ with $V_{\rm nsm}$ being the
effective potential on the new smooth path for 
global SUSY, which is constructed \cite{nsmooth} 
numerically.

\subsection{Inflationary observables}

We can make $S$ real by an appropriate global
${\rm U}(1)$ R transformation and define the
canonically normalized real inflaton field
$\sigma\equiv\sqrt{2}S$. The slow-roll parameters 
$\varepsilon$, $\eta$ and the parameter $\xi^2$, 
entering the running of $n_{\rm s}$, are given by 
(see e.g. Ref.~\cite{review})
\begin{equation}	
\varepsilon \equiv \frac{m_{\rm P}^2}{2}\,
\left(\frac{V^\prime(\sigma)}{V(\sigma)}\right)^2, 
\quad\eta \equiv m_{\rm P}^2\,\left(
\frac{V^{\prime\prime}(\sigma)}{V(\sigma)}\right),
\quad\xi^2 \equiv m_{\rm P}^4
\left(\frac{V^\prime(\sigma)V^{\prime\prime\prime}
(\sigma)}{V^2(\sigma)}\right),
\end{equation}	
where prime denotes derivation with respect to 
$\sigma$ and $V$ is either the effective potential
$V^{\rm SUGRA}_{\rm tr}$ on the trivial path given in 
Eq.~\eqref{Vtr}, or the effective potential 
$V^{\rm SUGRA}_{\rm nsm}$ on the new smooth path 
given in Eq.~\eqref{Vnsm}. 

Numerical simulations show that the system 
even during the intermediate phase follows basically 
the new smooth path. The e-foldings from the time when
the pivot scale $k_0=0.002~{\rm Mpc}^{-1}$ crosses
outside the horizon until the end of inflation are 
\cite{review}
\begin{equation}
N_Q\approx\frac{1}{m_{\rm P}^2}\,
\int_{\sigma_f}^{\,\sigma_c}\frac{V^{\rm SUGRA}_{\rm nsm}
(\sigma)}{V^{\rm SUGRA}_{\rm nsm}(\sigma)^\prime}\,d\sigma
+\frac{1}{m_{\rm P}^2}\,
\int_{\sigma_c}^{\sigma_Q}\frac{V^{\rm SUGRA}_{\rm tr}
(\sigma)}{V^{\rm SUGRA}_{\rm tr}(\sigma)^\prime}\,d\sigma,
\end{equation}
where $\sigma_Q\equiv\sqrt{2} S_Q>0$ is the value of 
$\sigma$ at horizon crossing of the pivot scale $k_0$ and 
$\sigma_f$ the value of $\sigma$ at the end of the second 
stage of inflation.

The power spectrum $P_{\mathcal R}$ of
curvature perturbation at $k_0$ is given \cite{review} by
\begin{equation}
P_{\mathcal R}^{1/2}\simeq
\frac{1}{2\pi\sqrt{3}}\,
\frac{V^{\rm SUGRA}_{\rm tr}(\sigma_Q)^{3/2}}
{m_{\rm P}^3V^{\rm SUGRA}_{\rm tr}(\sigma_Q)^\prime}.
\end{equation} 
The spectral index $n_{\rm s}$, the tensor-to-scalar
ratio $r$, and the running of the spectral index
$dn_{\rm s}/d\ln k$ are \cite{review}
\begin{equation}
n_{\rm s}\simeq 1+2\eta-6\varepsilon,\quad
r\simeq\,16\varepsilon, \quad
\frac{dn_{\rm s}}{d\ln k}\simeq 16\varepsilon\eta
-24\varepsilon^2-2\xi^2,
\end{equation}
with $\varepsilon$, $\eta$, and $\xi^2$ evaluated
at $\sigma=\sigma_Q$.
The number of e-foldings $N_Q$ required for solving 
the horizon and flatness problems is \cite{lectures}
\begin{equation}
N_Q\simeq53.76\,+\frac{2}{3}\,\ln\left(\frac{v_0}
{10^{15}~{\rm GeV}}\right)+\frac{1}{3}\,\ln
\left(\frac{T_{\rm r}}{10^9~{\rm GeV}}\right),
\end{equation}
where $T_{\rm r}$ is the reheat temperature and 
should be less than about $10^9~{\rm GeV}$ from the 
gravitino bound \cite{gravitino}.

\subsection{Monopole Production}

Magnetic monopoles are produced at the end of the 
standard hybrid stage of inflation, where
$G_{\rm PS}$ breaks to $G_{\rm SM}$, via the Kibble 
mechanism \cite{kibble}. The initial monopole number 
density is 
\begin{equation}
n_{\rm M}^{\rm init}\approx\frac{3{\sf p}}{4\pi}
H_0^3 ,
\end{equation}
since $H_0^{-1}$ is the relevant correlation 
length (${\sf p}\sim 1/10$ is a geometric factor). At 
the end of inflation, the monopole number density 
becomes
\begin{equation}
n_{\rm M}^{\rm fin}\approx\frac{3{\sf p}}{4\pi}
H_0^3e^{-3\delta N},
\end{equation}
where $\delta N$ is the number of e-foldings of the 
second inflationary stage.

Dividing $n_{\rm M}^{\rm fin}$ by the number density
$n_{\rm infl}\approx V^0_{\rm tr}/m_{\rm infl}$
of inflatons produced at the end of inflation
($m_{\rm infl}$ is the inflaton mass), we obtain 
\begin{equation}
\frac{n_{\rm M}}{n_{\rm infl}}\approx
\frac{3{\sf p}}{4\pi}H_0^3e^{-3\delta N}
\frac{m_{\rm infl}}{V^0_{\rm tr}}.
\end{equation}
This remains fixed until $T_{\rm r}$, where the 
relative monopole number density is \cite{thermal}
\begin{equation}
\frac{n_{\rm M}}{\sf s}=\frac{n_{\rm M}}
{n_{\rm infl}}\frac{n_{\rm infl}}{\sf s}\approx
\frac{3{\sf p}}{16\pi}\frac{H_0T_{\rm r}}
{m_{\rm P}^2}e^{-3\delta N}
\end{equation}
with ${\sf s}$ being the entropy density.

Taking $n_{\rm M}/{\sf s}\stackrel{_{<}}{_{\sim }}
10^{-30}$ (corresponding \cite{monreldens} to 
the Parker bound \cite{parker}), $T_{\rm r}\sim 
10^9~{\rm GeV}$, and $H_0\stackrel{_{<}}{_{\sim }}
10^{12}~{\rm GeV}$, we obtain
$\delta N\stackrel{_{>}}{_{\sim }} 9.2$, which 
implies that $N_{\rm st}\stackrel{_{<}}{_{\sim }}45$ 
($N_{\rm st}$ is the number of e-foldings of the 
pivot scale $k_0$ during standard HI). 
Here, $N_{\rm st}\ll 45$ and, thus, the present 
monopole flux is utterly negligible.

\subsection{Numerical Results}

\begin{figure}
\includegraphics[height=.5\textheight]{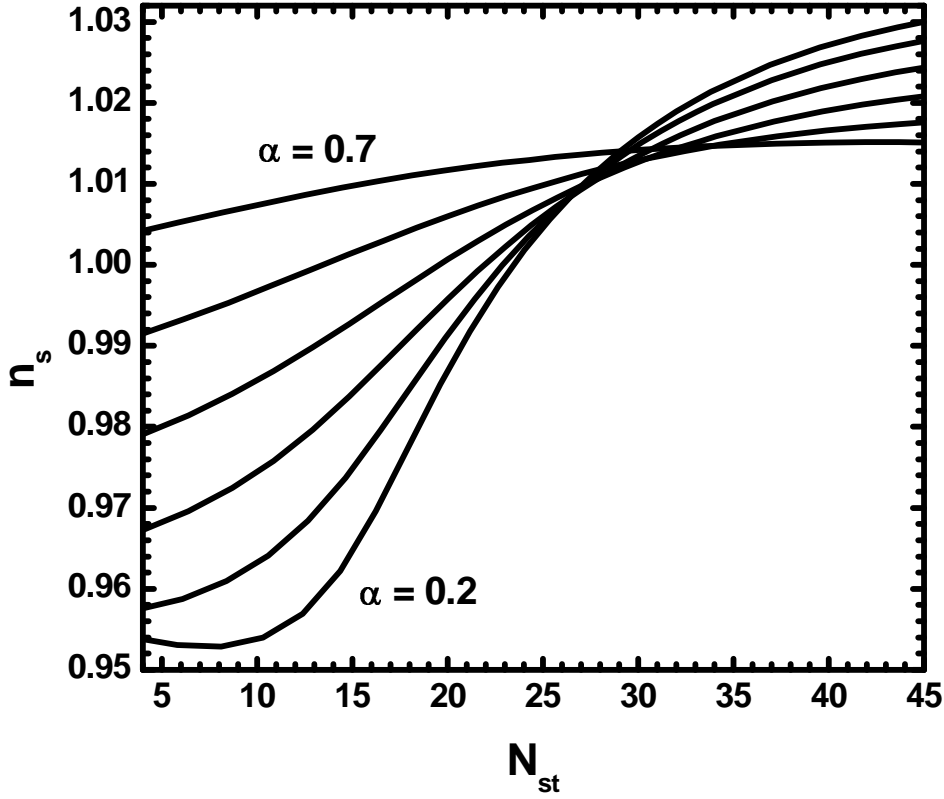}
\caption{Spectral index in standard-smooth HI 
versus $N_{\rm st}$ in minimal SUGRA
for $p=\sqrt{2}\kappa M/m=1/\sqrt{2}$. The values of
the parameter $\alpha$ range from $0.2$ to $0.7$
with steps of $0.1$.}
\label{nsSUGRAstsm}
\end{figure}

We put the mass $m_A$ of the color triplet,
anti-triplet gauge bosons divided by the gauge 
coupling constant $g\approx 0.7$ equal to the SUSY 
GUT scale $M_{\rm GUT}$ and the parameter 
$p=\sqrt{2}\kappa M/m$ equal to $1/\sqrt{2}$. Also, 
we take $T_{\rm r}\simeq 10^9~{\rm GeV}$ saturating 
the gravitino bound \cite{gravitino} and fix 
$P_{\mathcal R}^{1/2}\simeq 4.85\times 10^{-5}$ at 
the scale $k_0$ from the WMAP normalization \cite{wmap}. The 
resulting $n_{\rm s}$ is plotted \cite{sshi} against 
$N_{\rm st}$ and 
$\alpha=|\langle H^c\rangle|/|\langle\phi\rangle|$
in Fig.~\ref{nsSUGRAstsm}.

We took $4\stackrel{_{<}}{_{\sim }} N_{\rm st}
\stackrel{_{<}}{_{\sim }} 45$. The lower limit ensures 
that all the cosmological scales receive perturbations 
from the first stage of inflation, while the upper 
limit ensures that the present flux of monopoles in
our galaxy does not exceed the Parker bound \cite{parker}.
Note that, for $\alpha\stackrel{_{<}}{_{\sim }} 0.2$, 
$\lambda$ turns out to be non-perturbative, whereas, for
$\alpha\stackrel{_{>}}{_{\sim }} 0.7$, the WMAP 
normalization \cite{wmap} is not satisfied.

We see that $n_{\rm s}$'s below unity are readily obtainable 
and that the central value $n_{\rm s}=0.958$ from WMAP
is easily achievable, although spectral indices 
$n_{\rm s} \stackrel{_{<}}{_{\sim }} 0.953$ are not 
possible. We find that $n_{\rm s}$'s in the $95\%$ 
confidence level range \cite{wmap}
\begin{equation}
0.926 \stackrel{_{<}}{_{\sim }}n_{\rm s} 
\stackrel{_{<}}{_{\sim }} 0.99
\end{equation}
are obtained only if $N_{\rm st}\stackrel{_{<}}{_{\sim }} 
21$. So the present monopole flux is expected to be 
negligible.

The range of the various parameters of the model are 
\cite{sshi}
$\gamma\simeq(0.17-3.43)\times 10^{-3}$,
$\kappa\simeq(0.66-1.35)\times 10^{-2}$,
$\lambda\simeq 0.027-0.68$,
$M\simeq(2.12-2.44)\times 10^{16}~{\rm GeV}$,
$m\simeq(2.8-6.6)\times 10^{14}~{\rm GeV}$,
$\sigma_Q\simeq(0.95-3.05)\times 10^{17}~{\rm GeV}$,
$\sigma_c\simeq(0.6-2)\times 10^{17}~{\rm GeV}$,
$\sigma_f\simeq(4.9-9.9)\times 10^{16}~{\rm GeV}$,
$N_Q\simeq 54.1-54.5$,
$dn_{\rm s}/d\ln k\simeq-(0.77-3.76)\times 10^{-3}$,
and $r\simeq(0.7-5.3)\times 10^{-5}$.

\subsection{Gauge Unification}

Cosmology has constrained $m$ to be significantly 
lower than $M_{\rm GUT}$, which spoils gauge 
unification since some fields acquire masses 
$\sim m$. Actually, the fields with masses
$\stackrel{_{<}}{_{\sim }} M_{\rm GUT}$ turn out 
to be too many implying the existence of
Landau poles. Also, none of these fields has
$\rm SU(2)_L$ quantum numbers and, thus, the 
$\rm SU(2)_L$ gauge coupling constant fails to
unify with the other two gauge coupling constants.

Landau poles are avoided  by considering the 
superpotential term $\xi\phi^2\bar{\phi}$ allowed 
\cite{quasi} by all the symmetries of the model. 
This term 
gives masses $\sim|\xi\langle\phi\rangle|$ to some 
fields. The second problem is solved by including
a single extra superfield $\chi\in (15,3,1)$ with 
mass $m_{\chi}\approx 8\times 10^{14}~{\rm GeV}$ 
and charge $1/2$ under the $\rm U(1)$ R symmetry, 
which allows it to have just the superpotential 
term $m_{\chi}\chi^2/2$. Finally, for gauge 
unification, we find \cite{sshi} that 
$m\stackrel{_{>}}{_{\sim }} 4\times 
10^{14}~{\rm GeV}$, which implies that 
$\alpha\stackrel{_{<}}{_{\sim }} 0.5$.

\section{Conclusions}

The extension of the SUSY PS model introduced in 
Ref.~\cite{quasi} in order to solve the $m_b$ problem 
in SUSY GUTs with YU, such as the simplest SUSY PS 
model, universal boundary conditions and $\mu>0$ is 
a very fruitful framework for constructing HI
models. It naturally leads to new shifted and
new smooth HI using only renormalizable superpotential 
terms and avoiding the monopole problem. These 
variants, however, yield, in minimal SUGRA, too large 
$n_{\rm s}$'s.

This problem can also be resolved within the
same model by a two-stage inflationary scenario: 
the first stage is of the standard and the 
second of the new smooth hybrid type. Alternatively, 
we can have semi-shifted HI within the same model. This 
scenario incorporates cosmic strings contributing to 
the power spectrum of perturbations. In this case, 
larger $n_{\rm s}$'s are allowed.

\begin{theacknowledgments}
This work was supported by the European Union 
under the Marie Curie Initial Training Network 
"UNILHC" PITN-GA-2009-237920 and the Marie Curie 
Research Training Network "UniverseNet" 
MRTN-CT-2006-035863.
\end{theacknowledgments}

\bibliographystyle{aipproc}

\end{document}